\documentclass[12pt]{spieman}  % 12pt font required by SPIE;
\usepackage{amsmath,amsfonts,amssymb}
\usepackage{graphicx}
\usepackage{setspace}
\usepackage{tocloft}
\usepackage[left]{lineno} % hagino added

\title{Single Event Tolerance of X-ray SOI Pixel Sensors}

\author[a,*]{Kouichi~Hagino}
\author[b]{Mitsuki~Hayashida}
\author[b]{Takayoshi~Kohmura}
\author[b]{Toshiki~Doi}
\author[b]{Shun~Tsunomachi}
\author[b]{Masatoshi~Kitajima}
\author[c]{Takeshi~G.~Tsuru}
\author[c]{Hiroyuki~Uchida}
\author[c]{Kazuho~Kayama}
\author[d]{Koji~Mori}
\author[d]{Ayaki~Takeda}
\author[d]{Yusuke~Nishioka}
\author[d]{Masataka~Yukumoto}
\author[d]{Kira~Mieda}
\author[d]{Syuto~Yonemura}
\author[d]{Tatsunori~Ishida}
\author[e]{Takaaki~Tanaka}
\author[f]{Yasuo~Arai}
\author[g]{Ikuo~Kurachi}
\author[h]{Hisashi~Kitamura}
\author[i]{Shoji~Kawahito}
\author[i]{Keita~Yasutomi}
\affil[a]{Kanto Gakuin University, Research Advancement and Management Organization, 1-50-1 Mutsuura-higashi, Kanazawa-ku, Yokohama, Japan, 236-8501}
\affil[b]{Tokyo University of Science, School of Science and Technology, Department of Physics, 2641 Yamazaki, Noda, Chiba, Japan, 278-8510}
\affil[c]{Kyoto University, Faculty of Science, Department of Physics, Kitashirakawa-Oiwakecho, Sakyo-ku, Kyoto, Japan, 606-8502}
\affil[d]{University of Miyazaki, Faculty of Engineering, Department of Applied Physics, 1-1 Gakuen- Kibanadai-Nishi, Miyazaki, Miyazaki, Japan, 889-2192}
\affil[e]{Konan University, Department of Physics, 8-9-1 Okamoto, Higashinada, Kobe, Hyogo, Japan, 658- 8501}
\affil[f]{High Energy Accelerator Research Organization (KEK), Open Innovation Promotion Department, 1-1 Oho, Tsukuba, Ibaraki, Japan, 305-0801}
\affil[g]{D\&S Inc., 774-3-213 Higashiasakawacho, Hachioji, Tokyo, Japan, 193-0834}
\affil[h]{National Institute of Radiological Sciences, National Institutes for Quantum and Radiological Science and Technology, 4-9-1 Anagawa, Inage-ku, Chiba, Japan, 263-8555}
\affil[i]{Shizuoka University, Research Institute of Electronics, 3-5-1 Johoku, Naka-ku, Hamamatsu, Shizuoka, Japan, 432-8011}
\cftpagenumbersoff{figure}
\cftpagenumbersoff{table} 
\begin{document} 
\maketitle

%\linenumbers % hagino added

\begin{abstract}
We evaluate the single event tolerance of the X-ray { silicon-on-insulator (SOI)} pixel sensor named XRPIX, developed for the future X-ray astronomical satellite FORCE.
In this work, we measure the cross-section of { single event upset (SEU)} of the shift register on XRPIX by irradiating heavy ion beams with linear energy transfer (LET) ranging from $0.022{\rm ~MeV/(mg/cm^2)}$ to $68{\rm ~MeV/(mg/cm^2)}$.
From the SEU cross-section curve, the saturation cross-section and threshold LET are successfully obtained to be  { $3.4^{+2.9}_{-0.9}\times 10^{-10}{\rm ~cm^2/bit}$ and $7.3^{+1.9}_{-3.5}{\rm ~MeV/(mg/cm^2)}$}, respectively.
Using these values, the SEU rate in orbit is estimated to be $\lesssim 0.1{\rm ~event/year}$ primarily due to the secondary particles induced by cosmic-ray protons. This SEU rate of the shift register on XRPIX is negligible in the FORCE orbit.
\end{abstract}

% Include a list of up to six keywords after the abstract
\keywords{single event effect, silicon-on-insulator, X-ray detector, X-ray astronomy}

% Include email contact information for corresponding author
{\noindent \footnotesize\textbf{*}Kouichi Hagino,  \linkable{hagino@kanto-gakuin.ac.jp} }

%\begin{spacing}{2}   % use double spacing for rest of manuscript
\begin{spacing}{1}   % use double spacing for rest of manuscript

%%%%%%%%%%%%%%%%%%%%%%%%%%%%%%%%%%%%%
\section{Introduction}
\label{sec:introduction}
\begin{figure}[tb]
\centering
\includegraphics[width=0.6\hsize]{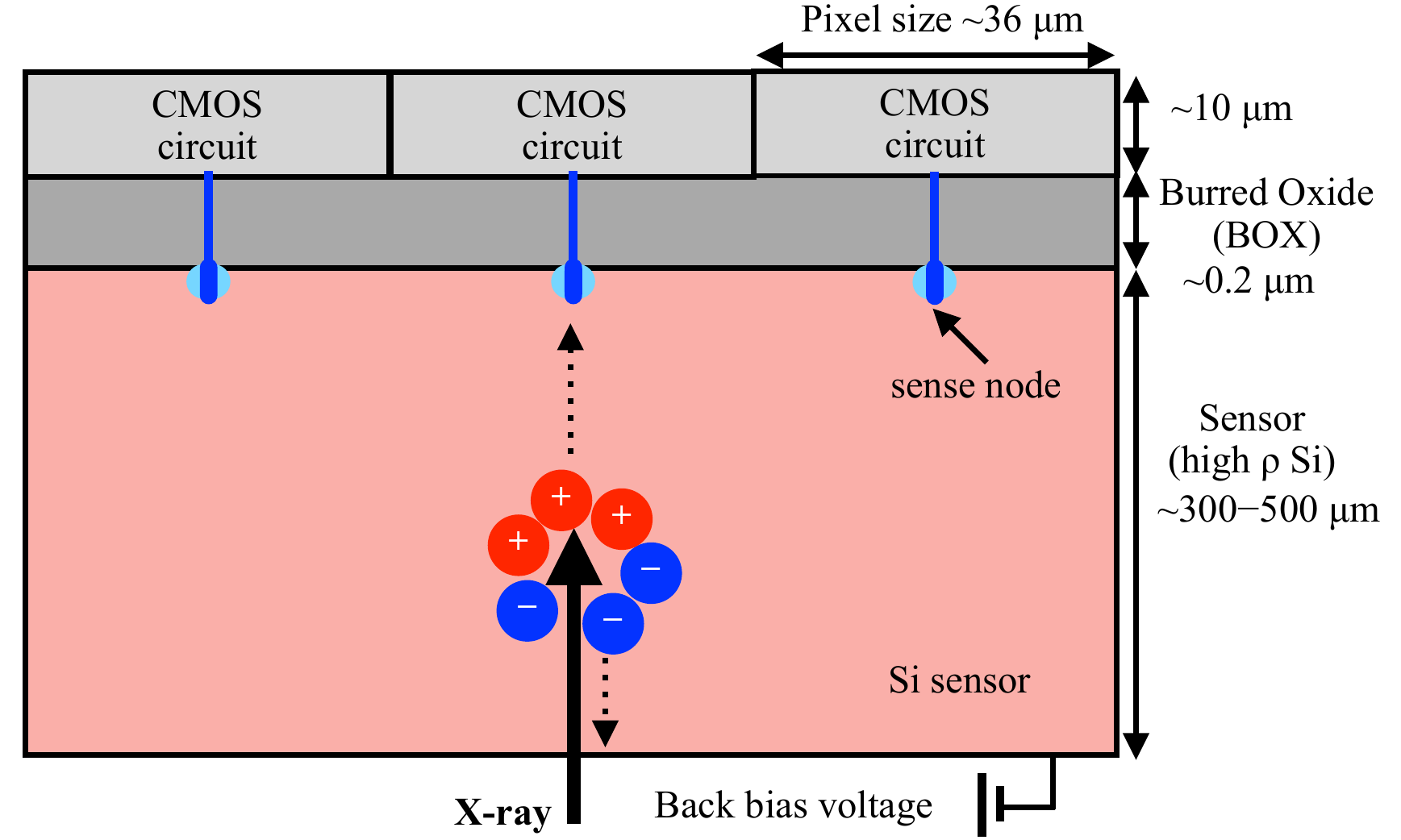}
\caption{Schematic view of XRPIX.}
\label{fig:XRPIX}
\end{figure}

The X-ray { silicon-on-insulator (SOI)} pixel sensor named XRPIX is a monolithic active pixel sensor developed as the main imaging spectrometer onboard the FORCE satellite\cite{Tsuru2018,Nakazawa2018,Mori2016}.
{ The FORCE satellite is an X-ray astronomical satellite aiming to be launched in the 2030s.
It will achieve a broadband X-ray imaging spectroscopy in an energy range of 1--79~keV with a high angular resolution better than 15~arcsec.
XRPIX is one of the main imagers of the FORCE satellite, and is fabricated with a $0.2{\rm ~\mu m}$ fully depleted SOI technology.}
Utilizing the SOI technology, XRPIX is composed of a high-resistivity Si sensor layer and a low-resistivity CMOS circuit layer with a SiO$_2$ insulator layer in between as shown in Fig.~\ref{fig:XRPIX}.
This structure enables to achieve a thick depletion layer as thick as a few hundred micrometers and to implement full CMOS readout circuitry on each pixel.
One of the most remarkable characteristics of XRPIX is the event-driven readout, where only signals from the pixels with X-ray events are readout, by using the trigger function implemented in each pixel circuit.
It realizes high timing resolution better than $\sim 10{\rm ~\mu s}$, which enables an extremely low background observation by adopting the anti-coincidence technique.

The tolerance to single event effect (SEE) should be considered in the development of the CMOS integrated circuits for space use. The SEE is a radiation effect caused by a strike of a single energetic particle. { There are two major categories in the SEE}: single event upset (SEU) and single event latch-up (SEL)~\cite{Velazco2007}.
In the SEU, the logical state of a digital circuit is changed by the free charge generated by the incident particle. It is not destructive but causes the malfunction of the circuit. On the other hand, the SEL is potentially destructive because it results in a large current by turning on a parasitic thyristor structure in the CMOS circuit.

Since the CMOS circuit of XRPIX is SOI-CMOS, single event tolerance should be better than that of bulk CMOS devices~\cite{Musseau1996}. In principle, the SEL cannot occur because there is no parasitic thyristor in the SOI-CMOS.
The SEU is also mitigated because the SEU-sensitive volume is reduced in the SOI-CMOS.
However, it is unknown whether the SEU tolerance of XRPIX is enough for the FORCE satellite.
In particular, in the FORCE satellite, the CMOS circuit of the flight model of XRPIX will contain a lot of shift registers storing operational parameters. If SEU will frequently occur in such shift registers, it would have a strong impact on the operations of XRPIX in orbit.
Thus, in this work, we measured the SEU cross-section curve of the shift register on XRPIX { for the first time}, and quantitatively evaluated its SEU tolerance.
The rest of the paper is organized as follows. In Sec.~\ref{sec:experiment} we describe the details of the heavy-ion irradiation experiment. In Sec.~\ref{sec:result} we present the main results, and estimate the SEU rate in orbit in Sec.~\ref{sec:estimation}. We conclude in Sec.~\ref{sec:conclusion}.

%%%%%%%%%%%%%%%%%%%%%%%%%%%%%%%%%%%%%
\section{Heavy-ion Irradiation Experiment}\label{sec:experiment}
\subsection{Test Device: XRPIX8}\label{sec:device}
We irradiated heavy ions to XRPIX at the Heavy Ion Medical Accelerator in Chiba (HIMAC) in the National Institute of Radiological Sciences.
The test device used in this experiment was the current prototype of the XRPIX series named ``XRPIX8''. XRPIX8 has a p-type sensor layer with a thickness of 300~${\rm \mu m}$. It has 96$\times$96 pixels with a pixel size of 36$\times$36~${\rm \mu m^2}$, so that the sensitive area is approximately $3.5\times 3.5{\rm ~mm^{2}}$. Adopting the pinned depleted diode (PDD) structure, XRPIX8 has an energy resolution as good as XRPIX6E, which achieved the best spectral performance among all the XRPIX series~\cite{Harada2018}.

In this experiment, we evaluated the single event tolerance of shift registers equipped in the on-chip peripheral circuits in XRPIX8~\cite{Takeda2020}.
The shift register is composed of a D-type flip-flop circuit, and used as a trigger mask for ignoring triggers from noisy rows or columns.
By writing a mask pattern to the shift register, trigger signals from the corresponding row or column are ignored.
Since XRPIX8 has $96\times 96$ pixels, there are two 96-bit shift registers for the trigger masks of rows and columns in XRPIX8.

In order to monitor the SEU in the shift registers, we periodically read out the values in the shift registers on XRPIX during the heavy-ion irradiation.
For each readout, SEU was judged by comparing the values in the shift register with the written values.
In addition to the SEU monitoring, we monitored the current consumption in XRPIX by applying voltages with source meters to detect SEL if it occurs.

\begin{figure}[tb]
\centering
\includegraphics[width=0.6\hsize]{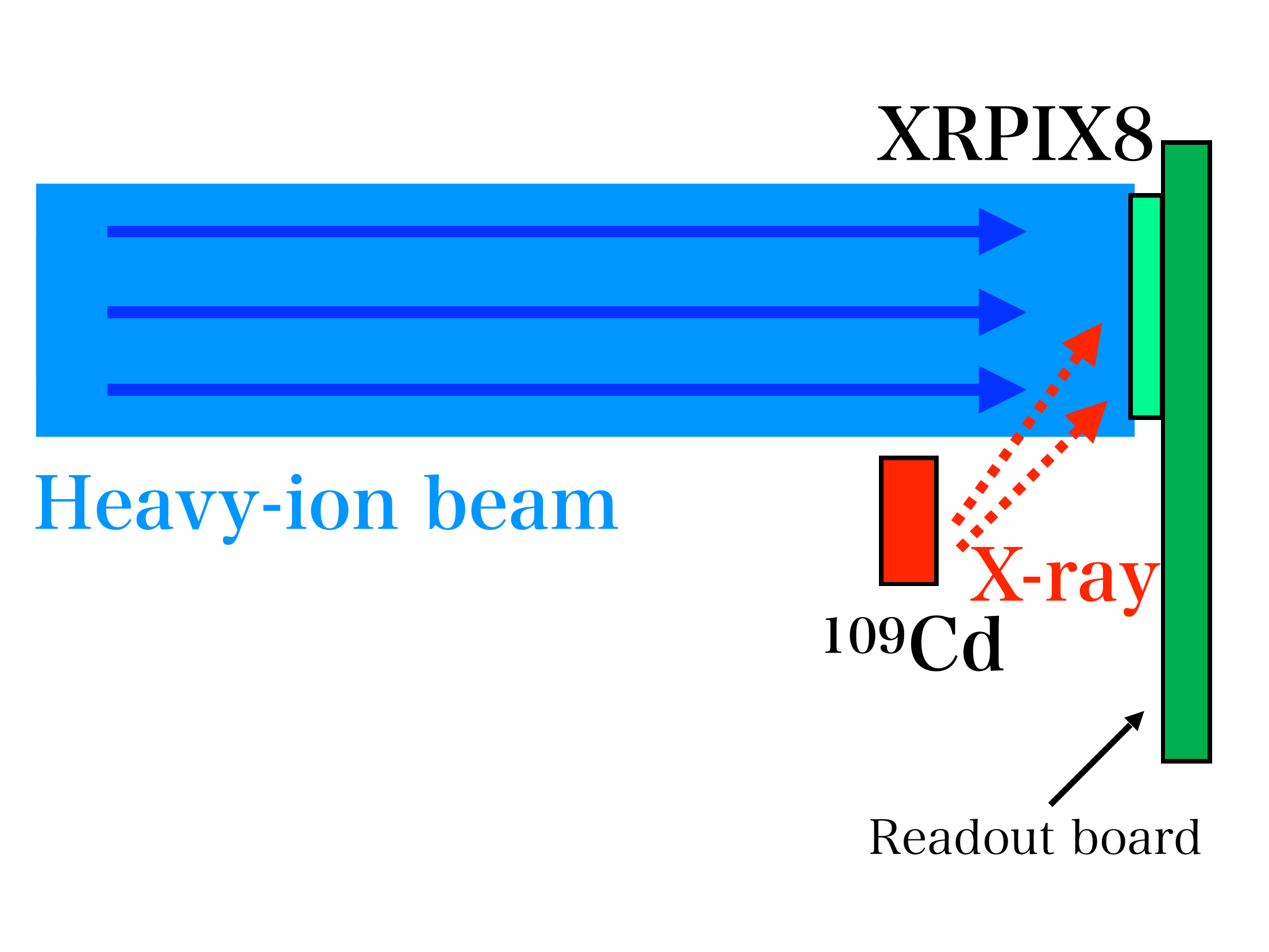}
\caption{Schematic picture of the experimental setup of heavy-ion irradiation of XRPIX.}
\label{fig:setup}
\end{figure}

During the heavy-ion irradiation, XRPIX was also irradiated with 22-keV X-rays from a radio isotope $^{109}$Cd as shown in Fig.~\ref{fig:setup}.
While the experiments with ions above 100~MeV/u were performed in the air, the experimental setup was located in a vacuum chamber for 6-MeV/u beams to avoid energy loss of the beam in the air.
The XRPIX sensor was operated under a back-bias voltage of $-25{\rm ~V}$, which creates approximately 100${\rm ~\mu m}$ of the depletion layer. Thus, both the heavy ions and X-rays were measured with XRPIX during the experiment.

\subsection{Heavy Ion Beam}
\begin{table}[tb]
\caption{Summary of the heavy-ion irradiation.} 
\label{tab:irradiation}
\begin{center}       
\begin{tabular}{|l |l |l |l |}
\hline
\rule[-1ex]{0pt}{3.5ex} 
Ion & Energy & LET & Fluence  \\
      & [MeV/u] & [${\rm MeV/(mg/cm^{2})}$] & [particles/cm$^{2}$]  \\
\hline\hline
\rule[-1ex]{0pt}{3.5ex} 
He & 100 & 0.022 & $2.96\times10^{9}$\\%3.63e8/(0.35*0.35)
\hline
\rule[-1ex]{0pt}{3.5ex} 
H & 6 & 0.051 & $8.65\times10^{9}$\\%1.06e9/(0.35*0.35)
\hline
\rule[-1ex]{0pt}{3.5ex} 
Si & 400 & 0.42 & $3.52\times10^{9}$\\%4.31e8/(0.35*0.35)
\hline
\rule[-1ex]{0pt}{3.5ex} 
Kr & 200 & 4.2 & $1.88\times10^{8}$\\%2.30e7/(0.35*0.35)
\hline
\rule[-1ex]{0pt}{3.5ex} 
Xe & 200 & 9.3 & $1.85\times10^{8}$\\%2.27e7/(0.35*0.35)
\hline
\rule[-1ex]{0pt}{3.5ex} 
Fe & 6 & 25 & $2.34\times10^{8}$\\%2.87e7/(0.35*0.35)
\hline
\rule[-1ex]{0pt}{3.5ex} 
Xe & 6 & 68 & $6.11\times10^{7}$\\%7.48e6/(0.35*0.35)
\hline
\end{tabular}
\end{center}
\end{table} 

To evaluate the SEU tolerance, it is necessary to determine the SEU cross-section as a function of linear energy transfer (LET).
Therefore, we adopted a variety of heavy ion beams with the LET values ranging from $0.022{\rm ~MeV/(mg/cm^2)}$ to $68{\rm ~MeV/(mg/cm^2)}$ as listed in Table~\ref{tab:irradiation}.
The LET values were estimated by using Geant4 simulation~\cite{Agostinelli2003,Allison2006,Allison2016}.
The total fluence at each LET was $6\times10^7$--$9\times10^9{\rm ~cm^{-2}}${, and the flux was $9\times10^{3}$--$2\times10^{6}{\rm ~cm^{-2}~s^{-1}}$}. Since the XRPIX device with PDD structure was established to work up to $\sim100{\rm ~krad}\simeq 6.3\times10^{9}{\rm ~MeV/mg}$ in our previous work~\cite{Hayashida2021}, we regulated the total fluence at each LET not to exceed it.

\begin{figure}[tb]
\centering
\includegraphics[width=0.8\hsize]{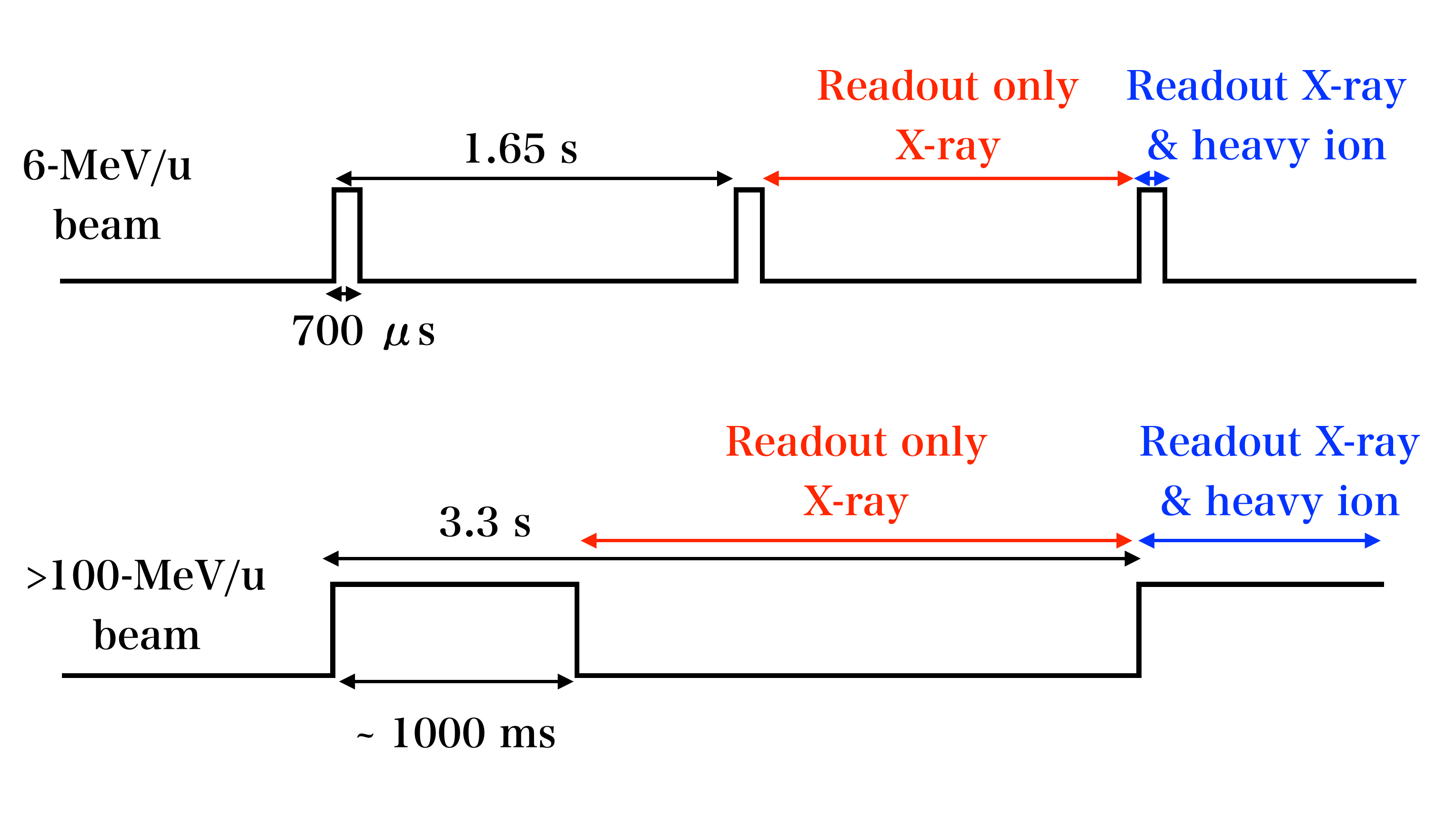}
\caption{Typical time profiles of the heavy-ion beams.}
\label{fig:beam_profile}
\end{figure}

Figure~\ref{fig:beam_profile} shows typical time profiles of the heavy-ion beams.
700-$\mu$s pulses were irradiated with a period of 1.65~s, while 1000-ms pulses (depending on ion) with a 3.3-s period.
Since X-ray is also irradiated as described in Sec.~\ref{sec:device}, only X-ray was readout during the beam-off phase, while both X-ray and heavy-ions were readout during the beam-on phase.

%%%%%%%%%%%%%%%%%%%%%%%%%%%%%%%%%%%%%
\section{Results of the Irradiation Experiment}\label{sec:result}
\subsection{Performance of XRPIX during the Heavy-ion Irradiation}
\begin{figure}[tb]
\centering
\includegraphics[width=0.8\hsize]{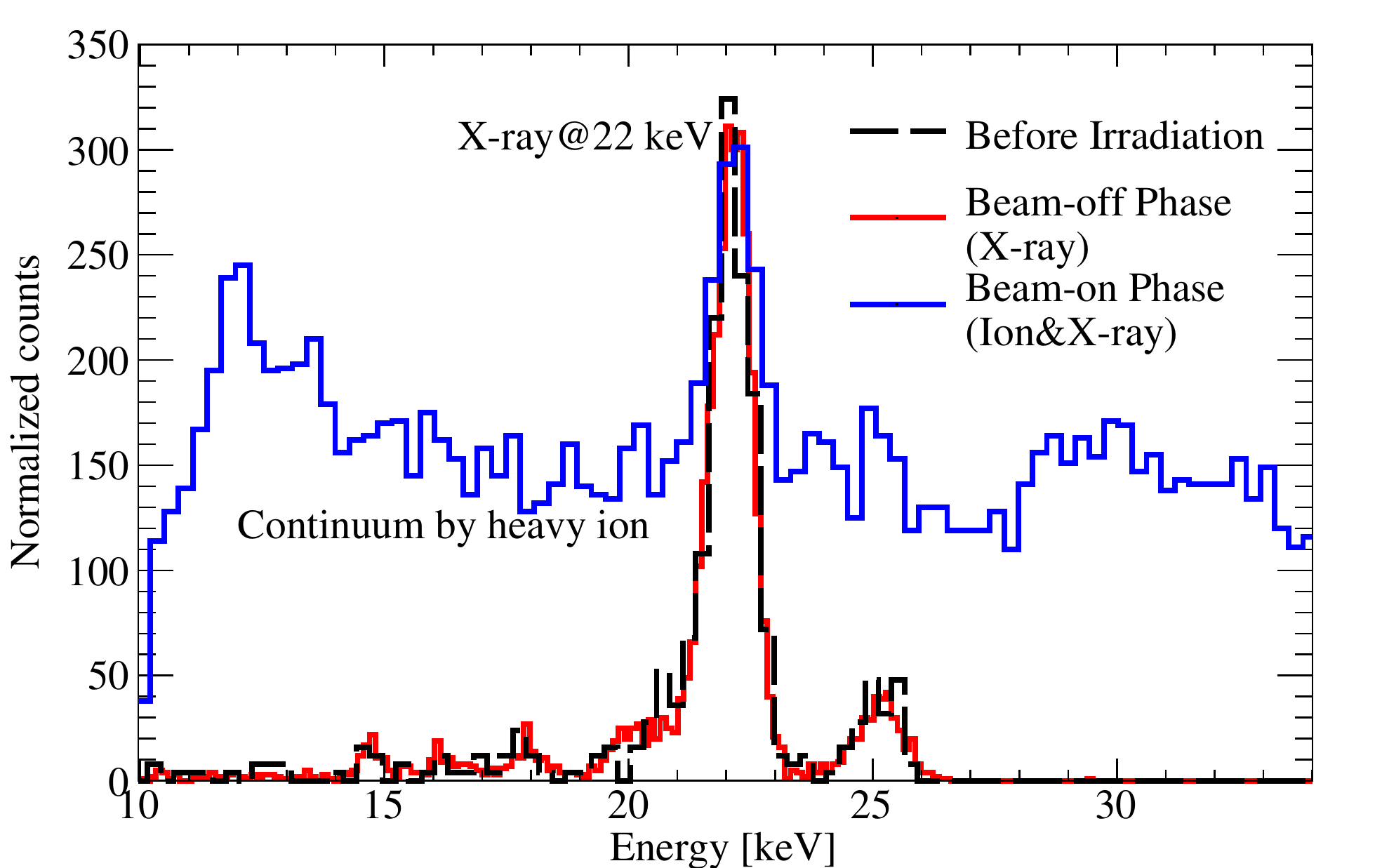}
\caption[$^{109}$Cd spectra measured with XRPIX during the heavy-ion irradiation.]{$^{109}$Cd spectra measured with XRPIX during the heavy-ion irradiation (red/blue) compared with that before the irradiation (black).}
\label{fig:spec}
\end{figure}

For the demonstration purpose, we show the measured spectra with XRPIX during the heavy-ion irradiation in Fig.~\ref{fig:spec}.
As shown in the figure, even in the beam-on phase, we were able to measure X-rays from $^{109}$Cd above the continuum by the heavy ions.
Also, in the beam-off phase, the spectral performance is almost the same as that before the irradiation.
Therefore, as demonstrated in Fig.~\ref{fig:spec}, XRPIX works in the event-driven readout mode without any major malfunctions even with the heavy-ion irradiation.

According to the current consumption monitoring of XRPIX during the irradiation, we did not find any signs of the SEL.
Throughout the experiment, the current consumption was kept at steady values.
It is a very plausible result because SEL never occurs in the SOI-CMOS devices in principle as described in Sec.~\ref{sec:introduction}.

\subsection{SEU Cross-section Curve}
\begin{figure}[tb]
\centering
\includegraphics[width=\hsize]{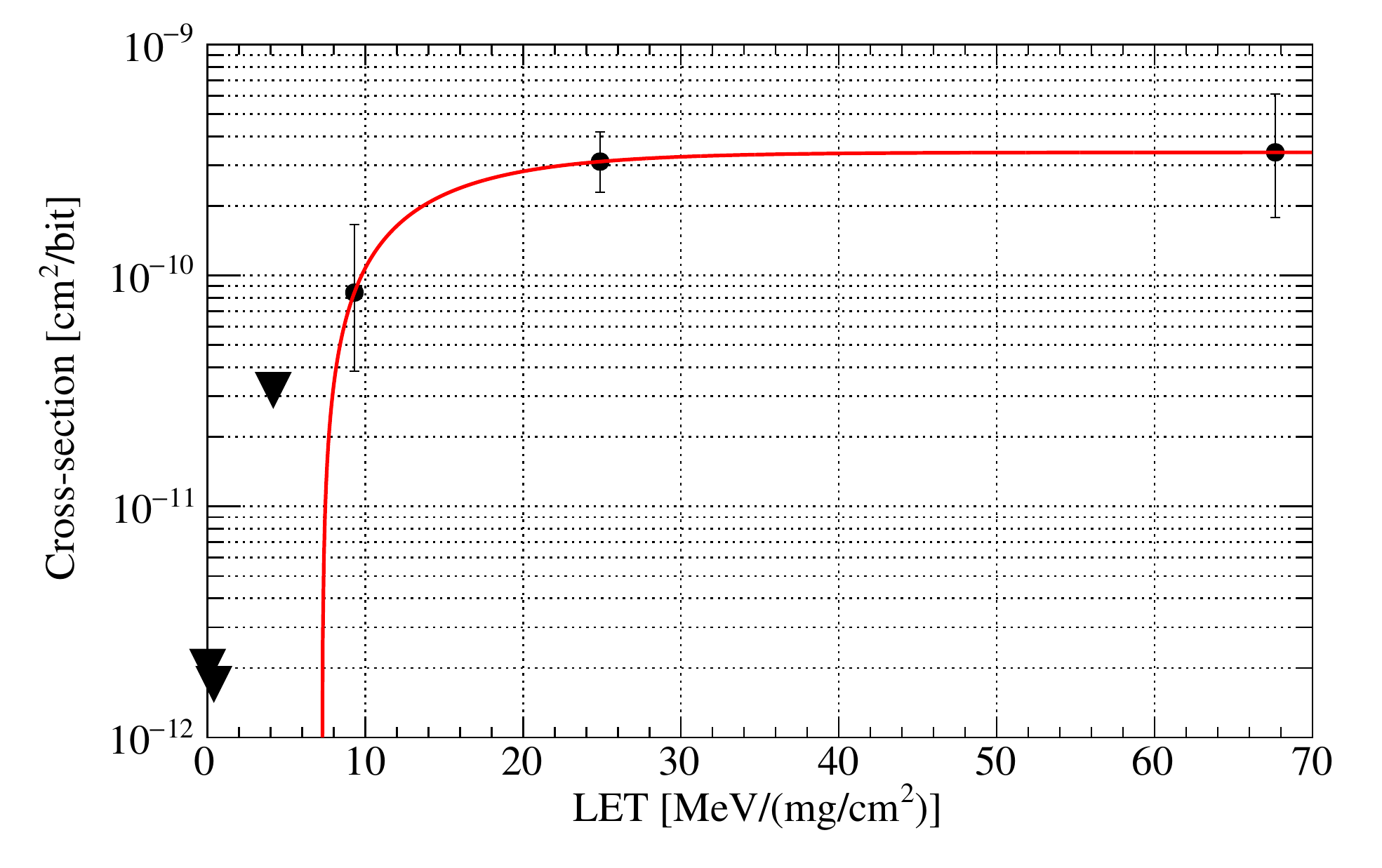}
\caption[SEU cross-section curve of the shift register on XRPIX and the best-fit Weibull function.]{SEU cross-section curve of the shift register on XRPIX (black points) and the best-fit Weibull function (red solid curve). Filled circles indicate the experimental data points with SEU detection, while triangles indicate 68\% upper limits without SEU detection.}
\label{fig:SEUcrosssection}
\end{figure}

Fig.~\ref{fig:SEUcrosssection} shows the SEU cross-section curve $\sigma(L)$ as a function of the LET value $L$ obtained by the heavy-ion irradiation experiment.
In general, the SEU cross-section curve is described as a Weibull function~\cite{Petersen1992}
\begin{equation}
\sigma(L)=\left\{
\begin{array}{ll}
\sigma_\infty\left[ 1-\exp\left(\frac{L-L_{\rm th}}{W}\right)\right] & (L \geq L_{\rm th})\\
0 & (L < L_{\rm th})
\end{array}
\right.,
\end{equation}
where $\sigma_{\infty}$ is the saturation cross-section, $L_{\rm th}$ is the threshold LET, and $W$ is the curve width.
Thus, by fitting with the Weibull function, we can obtain these characteristic parameters of the SEU cross-section curve.

Although we successfully detected the SEUs above $\sim9.5{\rm ~MeV/(mg/cm^2)}$, no SEU was detected below $\sim 4.5{\rm ~MeV/(mg/cm^2)}$, where only upper limits are shown as a triangle in the figure.
Thus, to utilize the data with no SEU detection, we adopted a maximum likelihood method of the binomial distribution for the fitting.
Since the SEU event is a Bernoulli trial where the stored bit is flipped or not, the number of SEU $k$ follows a binomial distribution
\begin{equation}
P(x)=\binom{n}{k}p^{k}(1-p)^{n-k},
\end{equation}
where $p$ is the probability of SEU and $n$ is the number of incidence particles.
Although it can be approximated to the Poisson distribution in our experiment, we implemented the likelihood function of the binomial distribution for wide applicability.
Thus, we minimized the likelihood function defined as
\begin{equation}
-2\ln L = -2\left[\ln \binom{n}{k} +k\ln p+(n-k)\ln(1-p)\right]
\end{equation}
for the SEU probability $p$ expected in the model, and obtained the best-fit parameters of the Weibull function.
We also estimated the uncertainties of parameters as the intervals where the difference of $-2\ln L$ from the best-fit was below unity.
As the result of fitting, we obtained the threshold LET of $L_{\rm th}=7.3^{+1.9}_{-3.5}{\rm ~MeV/(mg/cm^2)}$, the saturation cross-section of $\sigma_\infty=3.4^{+2.9}_{-0.9}\times 10^{-10}{\rm ~cm^2/bit}$, and the curve width of $W<35{\rm ~MeV/(mg/cm^2)}$ (68\% upper limit).

%%%%%%%%%%%%%%%%%%%%%%%%%%%%%%%%%%%%%
\section{Estimation of SEU Rate in Orbit}\label{sec:estimation}
Utilizing the obtained saturation cross-section and threshold LET, we estimated the SEU rate of the shift registers in XRPIX in the FORCE orbit.
{ We assumed radiation conditions in the solar maximum in this section.}
Since the flight model of XRPIX will have $\sim1000\times 1000$ pixels, the total number of bits in the shift registers for the trigger mask will increase to $\sim 2000$.
In addition, to reduce the size of the detector system, the flight model is planned to be equipped with digital circuitry, which is currently in the part of the readout board.
Although the number of bits implemented in the on-chip digital circuitry is not yet decided, it would be less than $10^{4}$ bits.
Thus, we used this conservative value as the total number of bits in the shift register to estimate the SEU rate.

We first estimated the SEU rate induced by heavy ions in orbit.
According to the Space Environment Information System {\it SPENVIS}~\cite{spenvis},  the heavy-ion flux above the threshold LET ($L_{\rm th}=7.3^{+1.9}_{-3.5}{\rm ~MeV/(mg/cm^2)}$) is $\lesssim 2\times 10^{-9}{\rm ~particles/cm^2/s}$ in the FORCE orbit. As the FORCE orbit, we assumed the low earth orbit with an altitude of 550~km and an inclination of 30$^{\circ}$~\cite{Mori2016}. By multiplying this flux, the assumed number of bits, and the measured cross-section, we estimated the SEU rate due to the heavy ions to be as small as $\sim 10^{-7} {\rm ~event/year}$.

In addition to the SEU by heavy ions, we also considered the SEU effect due to secondary particles generated by the incident cosmic-ray protons.
This effect has a significant contribution to the SEU in the FORCE orbit.
Although the cosmic-ray protons deposit small energies compared with the heavy ions, they can cause nuclear reactions with the device material.
This nuclear reaction generates the secondary particles, and they have large energy deposits on the device.
Thus, we need to estimate the SEU due to the secondary particles generated by the cosmic-ray protons.

According to the simple formula proposed in Barak~et~al.~\cite{Barak2006}, the SEU cross-section due to the proton-induced secondary particles is estimated to be $\sim10^{-15}{\rm ~cm^2/bit}$ at maximum.
Since the { flux of geomagnetically trapped protons} is $1\times10^2{\rm ~particle/cm^2/s}$ in the FORCE orbit, the proton-induced SEU rate is calculated to be $\lesssim 0.1{\rm ~event/year}$.
The actual SEU rate should be much smaller than this estimation in the FORCE satellite because XRPIX is planned to be surrounded by $\sim30$-mm thick BGO shields.
Therefore, the SEU rate of the shift registers on XRPIX is negligible in the FORCE satellite.

{
The SEU does not have a significant impact on the operation of XRPIX even if it occurs in orbit, though it was found to be a very rare event in this work.
Since the SEU is not a destructive event, if it occurs in space, it can be fixed by rewriting parameters to registers.
It could cause problems in celestial observation and satellite operation if the SEU frequently occurs.
However, in the case of low SEU probability as in this work, the effect of the SEU is avoidable by regularly rewriting the registers.
Therefore, we can conclude that the XRPIX has a sufficient SEU tolerance for the use onboard the FORCE satellite.
}

%%%%%%%%%%%%%%%%%%%%%%%%%%%%%%%%%%%%%
\section{Conclusions}\label{sec:conclusion}
We evaluated the SEU tolerance of XRPIX for the first time by irradiating the heavy-ion beams.
In this experiment, we found that XRPIX did not have any major malfunctions during the irradiation.
Also, we successfully estimated the threshold LET and saturation cross-section from the experimental results.
According to these values, the SEU rate of XRPIX in orbit is as rare as  $\lesssim 0.1{\rm ~event/year}$.
Therefore, we found that XRPIX has a sufficient SEU tolerance for the use onboard the FORCE satellite.

\subsection* {Acknowledgments}
We acknowledge the relevant advice and manufacture of the XRPIXs by the personnel of LAPIS Semiconductor Co., Ltd.
We also acknowledge Prof.~Munetaka~Ueno, Prof.~Masanobu~Ozaki, and Prof.~Hiroshi~Nakajima for the fruitful advice on the experimental planning.
This study was supported by JSPS KAKENHI Grant Numbers JP22H01269, JP25287042. This study was also supported by the VLSI Design and Education Center (VDEC), the University of Tokyo in collaboration with Cadence Design Systems, Inc., Mentor Graphics, Inc., and Synopsys, Inc.

%%%%% References %%%%%

\bibliography{report}   % bibliography data in report.bib

\begin{thebibliography}{10}

\bibitem{Tsuru2018}
T.~G. Tsuru, H.~Hayashi, K.~Tachibana, {\em et~al.}, ``{Kyoto's Event-Driven
  X-ray Astronomy SOI pixel sensor for the FORCE mission},'' in {\em
  Proceedings of SPIE},  A.~D. Holland and J.~Beletic, Eds.,  {\bf 10709}, 18,
  SPIE  (2018).

\bibitem{Nakazawa2018}
K.~Nakazawa, K.~Mori, T.~G. Tsuru, {\em et~al.}, ``{The FORCE mission: science
  aim and instrument parameter for broadband x-ray imaging spectroscopy with
  good angular resolution},'' in {\em Space Telescopes and Instrumentation
  2018: Ultraviolet to Gamma Ray},  J.-W.~A. den Herder, K.~Nakazawa, and
  S.~Nikzad, Eds.,  {\bf 48}, 84, SPIE  (2018).

\bibitem{Mori2016}
K.~Mori, T.~G. Tsuru, K.~Nakazawa, {\em et~al.}, ``{A broadband x-ray imaging
  spectroscopy with high-angular resolution: the FORCE mission},'' in {\em
  Proceedings of SPIE},  99051O  (2016).

\bibitem{Velazco2007}
R.~Velazco and F.~J. Franco, ``{Single Event Effects on Digital Integrated
  Circuits: Origins and Mitigation Techniques},'' in {\em 2007 IEEE
  International Symposium on Industrial Electronics},  3322--3327, IEEE
  (2007).

\bibitem{Musseau1996}
O.~Musseau, ``{Single-event effects in SOI technologies and devices},'' {\em
  IEEE Transactions on Nuclear Science} {\bf 43}, 603--613  (1996).

\bibitem{Harada2018}
S.~Harada, T.~G. Tsuru, T.~Tanaka, {\em et~al.}, ``{Performance of the
  Silicon-On-Insulator pixel sensor for X-ray astronomy, XRPIX6E, equipped with
  pinned depleted diode structure},'' {\em Nuclear Instruments and Methods in
  Physics Research Section A: Accelerators, Spectrometers, Detectors and
  Associated Equipment} {\bf 924}, 468--472  (2019).

\bibitem{Takeda2020}
A.~Takeda, K.~Mori, Y.~Nishioka, {\em et~al.}, ``{Development of on-chip
  pattern processing in event-driven SOI pixel detector for X-ray astronomy
  with background rejection purpose},'' {\em Journal of Instrumentation} {\bf
  15}, P12025--P12025  (2020).

\bibitem{Agostinelli2003}
S.~Agostinelli, J.~Allison, K.~Amako, {\em et~al.}, ``{Geant4―a simulation
  toolkit},'' {\em Nuclear Instruments and Methods in Physics Research Section
  A: Accelerators, Spectrometers, Detectors and Associated Equipment} {\bf
  506}, 250--303  (2003).

\bibitem{Allison2006}
J.~Allison, K.~Amako, J.~Apostolakis, {\em et~al.}, ``{Geant4 developments and
  applications},'' {\em IEEE Transactions on Nuclear Science} {\bf 53},
  270--278  (2006).

\bibitem{Allison2016}
J.~Allison, K.~Amako, J.~Apostolakis, {\em et~al.}, ``{Recent developments in
  Geant4},'' {\em Nuclear Instruments and Methods in Physics Research Section
  A: Accelerators, Spectrometers, Detectors and Associated Equipment} {\bf
  835}, 186--225  (2016).

\bibitem{Hayashida2021}
M.~Hayashida, K.~Hagino, T.~Kohmura, {\em et~al.}, ``{Proton radiation hardness
  of x-ray SOI pixel sensors with pinned depleted diode structure},'' {\em
  Journal of Astronomical Telescopes, Instruments, and Systems} {\bf 7}, 26
  (2021).

\bibitem{Petersen1992}
E.~Petersen, J.~Pickel, J.~Adams, {\em et~al.}, ``{Rate prediction for single
  event effects-a critique},'' {\em IEEE Transactions on Nuclear Science} {\bf
  39}, 1577--1599  (1992).

\bibitem{spenvis}
``{SPENVIS - Space Environment, Effects, and Education System}.''
  \url{https://www.spenvis.oma.be}.

\bibitem{Barak2006}
J.~Barak, ``{Simple Calculations of Proton SEU Cross Sections from Heavy Ion
  Cross Sections},'' {\em IEEE Transactions on Nuclear Science} {\bf 53},
  3336--3342  (2006).

\end{thebibliography}
\bibliographystyle{spiejour}   % makes bibtex use spiejour.bst

%%%%% Biographies of authors %%%%%

\vspace{2ex}\noindent\textbf{Kouichi Hagino} is an assistant professor at Kanto Gakuin University. He received his BS and MS degrees in physics from the University of Tokyo in 2010 and 2012, respectively, and his PhD degree in physics from the University of Tokyo in 2015. He has been working on the development of semiconductor detectors for applications in high-energy astrophysics.

\vspace{1ex}
\noindent Biographies and photographs of the other authors are not available.

\listoffigures
\listoftables

\end{spacing}
\end{document}